\documentclass[aps,prb,showpacs,byrevtex,showpacs,amsmath,amssymb,twocolumn]{revtex4}
\usepackage{amssymb}
\usepackage{epsfig}
\usepackage{graphicx}
\usepackage{color}

\newcommand{\srob}{Sr$_{3}$Ru$_{2}$O$_{7}$}
\newcommand{\sroa}{Sr$_{2}$RuO$_{4}$}

\begin{document}
\title{Structural and electronic properties of Sr$_{2}$RuO$_{4}$-Sr$_{3}$Ru$_{2}$O$_{7}$ heterostructure}
\author{Carmine Autieri}
\affiliation{SPIN-CNR, I-84084 Fisciano (Salerno), Italy}
\affiliation{Dipartimento di Fisica ``E.~R. Caianiello'',
Universit\`a di Salerno, I-84084 Fisciano (Salerno), Italy}
\affiliation{IAS, Forschungszentrum J$\ddot{u}$lich, 52425
J$\ddot{u}$lich, Germany}

\author{Mario Cuoco and Canio Noce}
\affiliation{SPIN-CNR, I-84084 Fisciano (Salerno), Italy}
\affiliation{Dipartimento di Fisica ``E.~R. Caianiello'', Universit\`a di
Salerno, I-84084 Fisciano (Salerno), Italy}

\begin{abstract}

By means of first-principles calculations we study the structural
and electronic properties of a superlattice made of Sr$_2$RuO$_4$
and Sr$_3$Ru$_2$O$_7$ ruthenate oxides. Due to the symmetry
mismatch of the two systems a significant structural rearrangement
occurs within the superlattice. We find that at the interface the
RuO$_6$ octahedra get elongated for the Sr$_2$RuO$_4$ while tend
to be compressed for the Sr$_3$Ru$_2$O$_7$ as compared to inner
layers and the bulk phases. The positions of the Sr-atoms in the
Sr-O layers at the interface are strongly modified and influence
the alignment of the Ru atoms with respect to the planar oxygens
as well as the Ru-O-Ru in-plane bond angles. Such structural
rearrangement leads to a modification of the electronic structure
close to the Fermi level. The main changes occurring at the
interface and in the inner layers of the heterostructure are
analyzed and compared with the bulk phases of the Sr$_2$RuO$_4$
and Sr$_3$Ru$_2$O$_7$ compounds. We show that the positions of the
peaks in the density of states close to the Fermi levels get
shifted and renormalized in the spectral weight. The balance
between the renormalization of the bandwidth of the d$_{xy}$ band
and that of the crystal field splitting results into a minor
change of the Van Hove singularities position within the
superlattice. The effective tight-binding parameters for the $4d$
Ru bands are determined by means of a maximally localized Wannier
functions approach and used to discuss the modification of the
electronic structure of the superlattice with respect to the bulk
phases. Consequences on the modification of the superconducting
and metamagnetic behaviour of the superlattice with respect to the
bulk phases are discussed.

\end{abstract}
\maketitle

\section{Introduction}

There is a consolidated evidence that the interface between
different electronic states and quantum orders is a source of
novel physical phenomena~\cite{Hwang12}. The interest for this
research area points both to the underlying fundamental physics as
well as to the high impact in applications based on
heterostructures with new emergent functionalities with respect to
their constituents. Transition metal oxides (TMO) with perovskite
structure are protoype systems to be exploited in this framework
due to the large variety of correlated driven physical phenomena
they exhibit, ranging from Mott insulator to unconventional
superconductivity through all sorts of different
spin-charge-orbital broken symmetry states~\cite{Imada98}. Such
aspect together with the recent achievements in the fabrication of
atomically controlled TMO-based interfaces explain why they
represent a unique laboratory to explore how spin, charge and
orbital reconstruction at the interface may determine novel
quantum states of matter
~\cite{Heber09,Ohtomo02,Gozar08,Ohtomo04,Nakagawa06,Tokura08,Dagotto07}.
On a general ground the reduced dimensionality at the interface is
certainly a key driving force as it may enhance the electronic
correlations against the kinetic energy. On the other hand, at a
microscopic level it is the degree of matching of the TMO forming
the heterostructure, the character of the transition elements and
how they get into contact at the interface to be a complex source
of a wide variety of physical properties.

In this context, the Ruddlesden-Popper (RP) family
Sr$_{n+1}$Ru$_n$O$_{3n+1}$ of Sr-based ruthenates, with $n$ being
the number of RuO layers in the unit cell, offers a distinct
perspective for designing interfaces of TMOs made of homologue
chemical elements but having a mismatch into the level of unit
cell complexity and in the character of the broken symmetry states
they exhibit as a function of $n$. The Sr-based RP family is made
of itinerant correlated systems with tendency to superconducting
or ferro- and meta-magnetic instabilities. The {\sroa} $n$=1
member is a spin-triplet superconductor with a topological
non-trivial chiral pairing.~\cite{Maeno94} The $n$=2 member,
Sr$_3$Ru$_2$O$_7$, is an enhanced Pauli paramagnet\cite{Ikeda00}
which at low-temperature exhibits an anisotropic metamagnetic
(i.e.\ field-induced) transition, with possible emergent nematic
states and unconventional quantum
criticality.\cite{Perry01,Grigera01} Finally, the $n$=3 member,
Sr$_4$Ru$_3$O$_{10}$, shows ferro- or meta-magnetic behavior
depending on the direction of the applied magnetic
field,\cite{Cao03,Mao06,Gupta06} and only SrRuO$_3$ ($n$=$\infty$)
is an isotropic ferromagnetic metal with $T_c\simeq160$
K.\cite{Allen96}

There are two reasons that make quite unique the use of ruthenates
RP-members in the interface design. First, the presence of an
eutectic point in the chemical phase diagram of the SrRuO
perovskites allows one to get natural interfaces in the form of
single crystalline micrometric domains between adjacent members of
the series, for instance Sr$_2$RuO$_4$-Sr$_3$Ru$_2$O$_7$ (i.e.
$n$=1 with $n$=2)~\cite{Fittipaldi05}. Interestingly, for the
eutectic systems interfaces between the two RP members may also
occur at the nanometer scale due to the intergrowth of the $n$=1
phase in the form of stacking faults layers in the $n$=2 domain
and viceversa~\cite{Fittipaldi05,Ciancio09}. By means of the same
growing technique, eutectic compounds of the type
Sr$_3$Ru$_2$O$_7$-Sr$_4$Ru$_3$O$_{10}$ (i.e. $n$=2 with $n$=3)
have been also achieved~\cite{Fittipaldi07} leading to interfaces
between unit cells mad of RuO bilayers and trilayers. The
investigation of the collective behavior of the eutectic phases
indicates a relevant role of the interface physics as the
superconducting~\cite{Hooper06,Fittipaldi08,Shiroka12} and the
magnetic properties turn out to be quite different if compared to
the homogeneous single crystalline ones. For instance, the
Sr$_2$RuO$_4$-Sr$_3$Ru$_2$O$_7$ eutectic shows an increase of the
superconducting transition temperature T$_c$~\cite{Shiroka12}
while for the Sr$_3$Ru$_2$O$_7$-Sr$_4$Ru$_3$O$_{10}$ there occurs
a shift in the metamagnetic critical field. Such observations seem
to point towards a significant electronic reconstruction at the
interface between the different members of the RP-series which in
turn results into a modification of the mechanisms responsible of
the observed quantum phases.

Another aspect of the research interest for interfaces made of
Sr-based ruthenates is that thin films of the series from $n$=1 to
$n$=5 \cite{Zurbuchen01,Tian07} have been successfully grown.
Thus, the synthesis of superlattices based on different RP-members
is an achievable task. In this framework, while the magnetic
states are usually obtained in the Sr-RP members in the shape of
thin films, it is quite remarkable to underline the recent
preparation~\cite{Krockenberger2010} of superconducting thin films
of Sr$_2$RuO$_4$ whose difficulty resides into the strong
sensitivity of the spin-triplet pairing to disorder. This
increases the research potential in designing superlattices or
heterostructures by fully employing the different character of the
broken symmetry states realized within the Sr-RP members as a
function of $n$.

\begin{figure}[t]
\centering
\includegraphics[width=0.45\textwidth, angle=0]{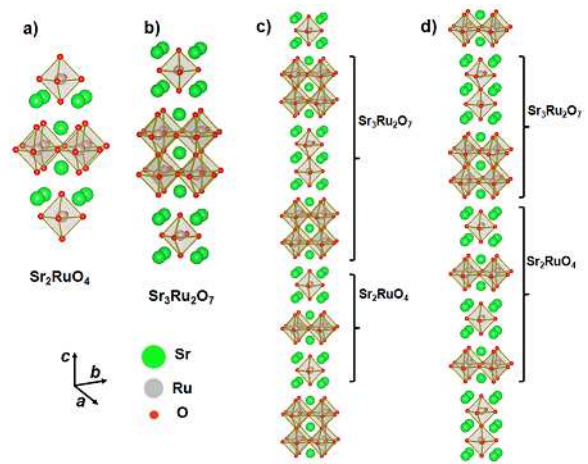}
\caption{Structure of the a) Sr$_2$RuO$_4$, b) Sr$_3$Ru$_2$O$_7$
bulk phases as well as of the c)
(Sr$_2$RuO$_4$)$_3$-(Sr$_3$Ru$_2$O$_7$)$_3$ and d)
(Sr$_2$RuO$_4$)$_4$-(Sr$_3$Ru$_2$O$_7$)$_2$ superlattices.}
\label{fig:strutt}
\end{figure}

Taking into account the above mentioned motivations, the aim of
the paper is to analyze the structural and electronic properties
of Sr$_2$RuO$_4$-Sr$_3$Ru$_2$O$_7$ superlattices. We use a
first-principles density functional theory to determine the fully
relaxed volume and the intra unit cell atomic positions as well as
the electronic structure of heterostructure made with $n$=1 and
$n$=2 unit cells. The Maximally Localized Wannier functions
approach is then applied to extract the effective tight-binding
parameters between the $4d$ Ru orbitals in the superlattice at the
interface and within the Sr$_2$RuO$_4$ and Sr$_3$Ru$_2$O$_7$ unit
cells. The outcome is used to discuss the modification of the
electronic structure of the superlattice, as well as its
interrelation with the structural properties and the bulk phases.
The focus is on two superlattice structures, as shown in Fig.
\ref{fig:strutt}(c)-(d), of the type
(Sr$_2$RuO$_4$)$_l$-(Sr$_3$Ru$_2$O$_7$)$_m$ with $(l,m)$=$(3,3)$ and
(4,2), $l$ and $m$ being the number of $n$=1 and $n$=2 unit cells
in the heterostructure, respectively. Such configurations consist
of at least an interface layers block that is inequivalent to the
inner one within the $n$=1 or the $n$=2 side of the
heterostructure. This allows one to extract the structural and the
electronic features of the different Ru-O layers depending on the
character of the neighboring ones. Due to the different symmetry
and size of the $n$=1 and $n$=2 unit cells a rearrangement of the
atomic position takes place both within the Ru-O and Sr-O layers.
We show that the RuO$_6$ octahedra at the
Sr$_2$RuO$_4$-Sr$_3$Ru$_2$O$_7$ interface get elongated in the
Sr$_2$RuO$_4$ side and compressed in the Sr$_3$Ru$_2$O$_7$ along
the $c$-axis when compared to the inner layers block and to the
bulk phases as well. The positions of the Sr atoms in the Sr-O
layers get modified in a way to increase the Ru-O misalignment in
the $ab$ plane and result into an asymmetric planar bond length
with respect to that with the apical oxygen. The interface layers
undergo the major structural rearrangement which in turn
influences the electronic structure close to the Fermi energy.
Though the superlattice configuration would point to a greater
change in the bands with d$_{xz}$ and d$_{yz}$ symmetry that
mainly overlap along the $c-$axis, interesting variations are also
observed for the d$_{xy}$ band as due to the induced rotation of
the RuO$_6$ octahedra and the modification of the crystal field
splitting. We show that the van Hove singularities of the
Sr$_2$RuO$_4$ and Sr$_3$Ru$_2$O$_7$ bulk systems get shifted in
energy in the superlattice and that such modification is more
pronounced at the interface layers. As a signature of the
hybridization between the $n$=1 and $n$=2 bands, there is a
spectral weight enhancement for the d$_{xy}$ band at energies
which roughly correspond to those of the van Hove position in the
Sr$_3$Ru$_2$O$_7$.

The paper is organized as follows. In the Sect. II we provide the
general framework of the computational analysis. Sect. III is
devoted to the presentation of the results concerning the
structural and the electronic outcome for the Sr$_2$RuO$_4$,
Sr$_3$Ru$_2$O$_7$ bulk phases as well as for the
(Sr$_2$RuO$_4$)$_l$-(Sr$_3$Ru$_2$O$_7$)$_m$ superlattices. In the
Sect. IV we present the concluding remarks.

\section{Computational framework}

We perform unpolarized first-principles density functional
calculations~\cite{Kohn64} by using the plane wave ABINIT
package,~\cite{Gonze05} the Generalized Gradient Approximation
(GGA) for the exchange-correlation functional,~\cite{Wu06} and
ultrasoft pseudopotentials.~\cite{Fuchs99} We consider a
plane-wave energy cut-off of 80 Ry and a cold smearing of 0.045
Ry. These values are common to all the presented calculations. The
strategy we apply is to first perform the calculation of the
relaxed crystal structure and of the energy spectra for the
{\sroa} and {\srob} bulk phases and then to implement the same
computational scheme for the superlattice. Since similar
calculations have been already performed and a lot of data are
available on the structural and electronic properties of the
{\sroa} and {\srob} bulk phases, we have profited of such know-how
to test the accuracy of the used approach. In this way, we feel
confident to have a controlled scheme of analysis for the
homogeneous and the superlattice structures. We would like to
notice that standard functionals based on local density
approximation may tend to overestimate the volume producing
inaccurate ratio between the $c$- and $a$-axes.~\cite{Bilic08}
This problem is encountered for the analyzed class of ruthenate
oxides, especially for the Sr$_3$Ru$_2$O$_7$ bulk phase. Thus, to
get a more accurate determination of the volume and lattice
constants in the study of the superlattice made of the first two
RP members of the series, we employ the exchange-correlation of Wu
and Cohen,~\cite{Wu06} a variant of the generalized gradient
approximation by Perdew {\it et al} (PBE)~\cite{Perdew96} method
optimized for the relaxation of bulk systems. An $8\times 8\times
8$ k-point grid is used for {\sroa}, while a $8\times 8\times 2$
grid used for {\srob} for the full relaxation. Furthermore, the
Sr$_2$RuO$_4$-Sr$_3$Ru$_2$O$_7$ hybrid structures are studied with
a $4\times 4\times 1$ k-point grid for the full relaxation and a
$8\times 8\times 1$ k-point grid for the determination of the
density of states (DOS). We optimize the internal degrees of
freedom by minimizing the total energy to be less than $10^{-8}$
Hartree and the remaining forces to be less than $10^{-4}$
Hartree/Bohr, and we require the external pressure to be less than
0.05 GPa to obtain the full relaxation of the system.

Let us consider some details about the determination of the
effective tight-binding Hamiltonian in an atomiclike Wannier basis
for the {\sroa}, {\srob} and the superlattice configurations.
There are different ways to get the Wannier functions for the
relevant electronic degrees of freedom including the
orthogonalized projections of specific atomic orbitals on the
Bloch wave-functions in a distinct energy window and downfolded
muffin tin orbitals as well as maximally localized Wannier
functions (MLFW). Hereafter, to extract the character of the
electronic bands at the Fermi level, we use the Slater-Koster
interpolation scheme based on the Maximally Localized Wannier
Functions method (MLWF). Such approach is applied to the
determination of the real-space Hamiltonian matrix elements in the
MLWF basis for the bulk Sr$_2$RuO$_4$ and Sr$_3$Ru$_2$O$_7$ phases
as well as for the different superlattice configurations and to
discuss the modification of the relevant parameters with respect
to the structural changes. After obtaining the DFT Bloch bands
within GGA, the MLWFs are constructed using the WANNIER90 code~\cite{Mostofi08}.
Starting from an initial
projection of atomic $d$ basis functions belonging to the t$_{2g}$
sector and centered at the different Ru sites within the unit cell
on the Bloch bands we get a set of three t$_{2g}$-like MLWF for
each site within the different unit cells of the analyzed systems.


Then, we introduce the main
concepts about the MLWFs procedure. We start
by noticing that in general the Block waves can be expressed as a
Bloch sum of atomiclike basis functions or Wannier functions.
Indeed, assuming to have a group of $N$ Bloch states
$|\psi_{nk}\rangle$ that is isolated in energy from the other
bands in the Brillouin zone (BZ), one can construct a set of $N$
localized Wannier functions $|w_{n \bf{R}}\rangle$ associated with
a lattice vector $\bf{R}$ by means of the following transformation:
\begin{eqnarray*}
|w_{n{\bf R}}\rangle =\frac{V}{(2\pi)^3} \int_{BZ}
\left(\sum_{m=1}^{N} U^{(\bf{k})}_{mn} |\psi_{n \bf{k}}\rangle
\right) e^{-i\,\mathbf{k}\cdot \mathbf{R}} d{\mathbf{k}}
\end{eqnarray*}
where, $U^{(\bf{k})}$ is a unitary matrix that mixes the Bloch
functions at a given $\bf{k}$-vector in the Brillouin zone. The
choice of $U^{(\bf{k})}$ determines the structure of the Wannier
orbitals. In Ref. \cite{Marzari97} the authors demonstrated that a
unique set of Wannier functions can be obtained by minimizing the
total quadratic spread of the Wannier orbitals expressed in terms
of the position operator $r$ through the following relation
$\Omega=\sum_{n=1}^{N} \left[ \langle r^2\rangle_n- \langle r
\rangle_n{^2} \right]$ with $\langle O \rangle_n=\langle w_{n0}|
O|w_{n0} \rangle$. For the case of entangled bands, one has to
introduce another unitary matrix that takes into account the extra
Block bands in the energy window upon examination. Such matrix is
also obtained by minimizing the functional $\Omega$.\cite{Souza01}
Once a set of MLWFs is determined the corresponding matrix
Hamiltonian is given by a unitary transformation from the diagonal
one in the Bloch basis. The resulting real space representation of
the Hamiltonian in the MLWF basis can be expressed as
\begin{eqnarray*}
\tilde{H}=\sum_{{\bf R},\bf{d}} t^{{\bf d}}_{nm}
\left(\tilde{c}^{\dagger}_{n,{\bf{R}}+{\bf{d}}}
\tilde{c}_{m,\bf{R}} +h.c. \right)
\end{eqnarray*}
here $\tilde{c}_{m,\bf{R}}$ destroy an electron in the $n$ orbital
Wannier state $|w_{n \bf{R}}\rangle$. Then, the real space
elements $t^{{\bf d}}_{nm}$ can be considered effective hopping
amplitudes as in a tight-binding approach between MLWF separated
by a distance $\bf{d}$ associated with the lattice vectors. The
index $n$ refers to site and orbitals in the case of a unit cell
that contains more than one site.


\section{Results}

In this section we present the structural and electronic
properties of the Sr$_2$RuO$_4$-Sr$_3$Ru$_2$O$_7$ superlattice.
Before considering the hybrid structure, we analyze the electronic
structure of the bulk Sr$_2$RuO$_4$ and Sr$_3$Ru$_2$O$_7$ phases
within the computation scheme above mentioned. We remind that
Sr$_2$RuO$_4$ has a space group $I4/mmm$, whereas, due the
rotations of the octahedra, Sr$_3$Ru$_2$O$_7$ presents an
orthorhombic symmetry with $Pban$ space group.

\subsubsection{Bulk {\sroa} phase}

The electronic structure of Sr$_2$RuO$_4$ has been already studied
by several authors by means of density functional theory with
Local Density Approximation (LDA) or
GGA.~\cite{Oguchi95,Singh95,Mazin99,Liebsch00,Pavarini06} The
tetravalent Ru atom has four electrons left in the $4d$ shell; the
quasi-cubic crystal field splits the $d$ levels into three-fold
degenerate $t_{2g}$ and two-fold degenerate $e_g$ states. At the
Fermi level, four electrons fill the $t_{2g}$ bands, while the
$e_g$ bands are empty, being higher in energy. DFT calculations
show that the three $t_{2g}$ bands can be divided into a wider
quasi-two-dimensional $xy$, and two quasi-one-dimensional bands of
$\gamma z$ character. While the first band does not hybridize with
the others, the $xz$ and $yz$ mix each other by forming an
electron and hole-like Fermi sheets.

\begin{figure}[t]
\centering
\includegraphics[width=0.5\textwidth, angle=0]{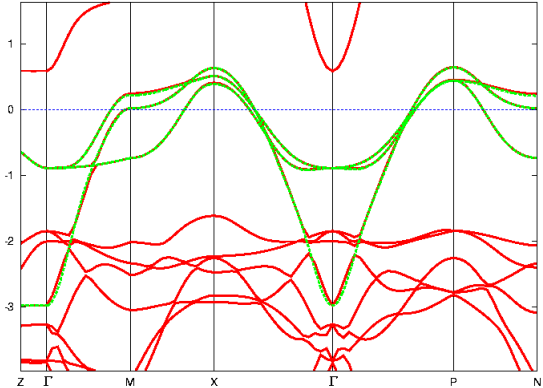}
\caption{Electronic structure of Sr$_2$RuO$_4$ obtained within the
density functional analysis by means of GGA (red full line) and
the dispersion obtained from the t$_{2g}$-like MLWFs (green dotted
line). The $e_g$ levels are about 0.5 eV above the Fermi level.
The Fermi level is set to zero and the energies are reported in
eV.} \label{fig:BS_214}
\end{figure}

The electronic structure and the dispersion calculated from the
t$_{2g}$-like MLWFs are shown in Fig.~\ref{fig:BS_214}. The
spectra and the derived Fermi surface is in agreement with those
of the previous studies. As a distinct feature that will be also
analyzed later on for the superlattice we notice that the energy
spectrum exhibits a flatness at the $M=(0,\pi)$ and $(\pi,0)$
points of the Brillouin zone leading to a van Hove singularity in
the density of states above the Fermi level. The MLWFs approach
provides the effective electronic parameters for the
$t_{2g}$-atomiclike obtained Wannier functions. The resulting
amplitudes are reported in the Table \ref{Tab1:214} along specific
connecting vectors in the $ab$ plane and along the $c$-axis. The
amplitude of the effective electronic parameters is in agreement
with those deduced by means of different first principles
methods.\cite{Pavarini06}

We point out that DFT calculations based on
LDA~\cite{Oguchi95,Singh95,Mazin99} place the ($\pi$, 0), (0,
$\pi$) saddle point of the $xy$ band about 60~meV above the Fermi
energy. By taking into account the gradient corrections this
singularity is lowered in energy to about 50~meV above the Fermi
level,~\cite{Mazin99} while local Coulomb correlations push the
$xy$ VHS approximately to within 10~meV, near the Fermi
energy.~\cite{Liebsch00,Pavarini06}

Concerning the structural properties, the relaxed in-plane
(out-of-plane) lattice constant $a$ ($c$) turns out to be slightly
larger (smaller) than the available experimental value with a
resulting computed volume which is only $\sim$0.7\% larger than
the experimental one.
For completeness, it is worth pointing out that the volume
obtained using PBE is $\sim$1\% larger than experimental
value.~\cite{Matzdorf00}


%
%

\small
\begin{table*}
\caption{Hopping integrals along the direction $[lmn]$ and on-site
energy in eV associated to the three orbitals of the t$_{2g}$
sector of the bulk \sroa\ at experimental atomic
positions~\cite{Chmaissen98}. The connecting vector is expressed
in terms of the integer set $[l\,m\,n]$ and the lattice constants
$a$ and $c$ as
$\textbf{d}=l\,a\,\textbf{x}+m\,a\,\textbf{y}+n\,c\,\textbf{z}$.~\cite{Gonze05,Gonze09,Mostofi08}
} \small
\begin{center}
\begin{tabular}{|l|l|l|l|l|l|l|l|l|}
\hline
\multicolumn{1}{|l|}{orbital index} & \multicolumn{8}{c|}{amplitude} \\
\hline $[l\,m\,n]$ & [000] & [100] & [010] & [110] & [200] & [020]
& [$\frac{1}{2}$\,$\frac{1}{2}$\,$\frac{1}{2}$]
 & [001] \\
 xy-xy & -0.4750 & -0.3867 & -0.3867 & -0.1384 & 0.0094 & 0.0094 & 0.0017 & -0.0013\\
 yz-xy & 0 & 0 & 0 & 0 & 0 & 0 & 0.0057 & 0 \\
 xz-xy & 0 & 0 & 0 & 0 & 0 & 0 & 0.0057 & 0 \\
 yz-yz & -0.3224 & -0.0389 & -0.2914 & 0.0165 & 0.0010 & 0.0612 &
 -0.0188 & 0.0006 \\
yz-xz & 0 & 0 & 0 & -0.0121 & 0 & 0 & -0.0136 & 0 \\
xz-xz & -0.3224 & -0.2914 & -0.0389 & 0.0165 & 0.0612 & 0.0010 &
-0.0188 & 0.0006 \\
\hline
\end{tabular}
\end{center}
\label{Tab1:214} \normalsize
\end{table*}
\normalsize

\subsubsection{Bulk {\srob} phase}

The electronic structure of Sr$_3$Ru$_2$O$_7$ has been analyzed
using ARPES\cite{Puchkov98,Tamai08} and density functional
calculations\cite{Singh01} with a reasonable agreement between
theory and experiments. Due to the presence of two RuO layers in
the unit cell, as compared with the Sr$_2$RuO$_4$ case, there are
replica of the $t_{2g}$ bands which are then split by the bilayer
coupling and the orthorhombic distortions. To a first
approximation the Fermi surface of the Sr$_3$Ru$_2$O$_7$ can be
derived from the six $t_{2g}$ bands (three from each RuO layer)
with bonding-antibonding (odd-even) splitting due to the bilayer
coupling. Nevertheless the topology of the Fermi surface is deeply
modified with respect to a simple doubling of the Fermi sheets due
to various factors. The bilayer splitting, the extra $k_z$
dispersions, and the avoiding crossing between the d$_{xy}$ and
the d$_{\gamma z}$ bands tend to reduce the bands nesting.
Furthermore, due to the orthorhombic distortions, small
cylindrical lens shaped Fermi sheets form around the M point (i.e.
the midpoint of the original $\Gamma$ to $X$ direction in the
tetragonal Brillouin zone) and two small cylindrical sheets with
almost no $k_z$ dispersion around the $\Gamma$ point.

\begin{figure}[!]
\centering
\includegraphics[width=0.3\textwidth, angle=270]{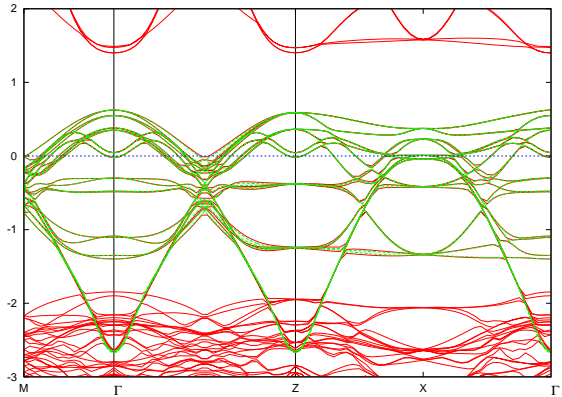}
\caption{Electronic structure of Sr$_3$Ru$_2$O$_7$ obtained within the
density functional analysis by means of GGA (red full line) and
the dispersion obtained from the t$_{2g}$-like MLWFs (green dotted
line). The position of the Van Hove like singularity is located in
the middle of the long direction Z-$\Gamma$ (X point). The position of the VHS is the same of  Sr$_2$RuO$_4$, but, the crystal symmetry is different. The Fermi
level is set to zero and the unit scale of the energies is eV.}
\label{fig:BS_327}
\end{figure}

We have performed a detailed analysis of the electronic structure
of the bilayer Sr$_3$Ru$_2$O$_7$ within the GGA scheme
described in the Sect. II by focusing on the case of the fully
distorted orthorhombic configuration.~\cite{Huang98} The obtained
band structure reported in Fig.~\ref{fig:BS_327} as well as the
derived Fermi surface fairly agree with the ARPES and the previous
density functional results above mentioned. Due to the various
physical factors that enter into the electronic structure and the
difficulty to trace and correlate them when a superlattice
configurations is considered, we have determined the effective
tight-binding parameters in the MLWFs basis.

\small
\begin{table*}
\caption{Effective hopping parameters along the direction $[lmn]$
and on-site energy in eV associated to the t$_{2g}$-like Wannier
functions for the {\srob}. The connecting vector is expressed in
terms of the integer set $[l\,m\,n]$ and the lattice constants $a$
and $c$ as
$\textbf{d}=l\,a\,\textbf{x}+m\,a\,\textbf{y}+n\,c\,\textbf{z}$
~\cite{Gonze05,Gonze09,Mostofi08}. The direction $00p$ connect the
two ruthenium atoms in the bilayer. The hopping parameters that
are zero in the tetragonal phase have sign dependent from the
tilting (clockwise or anticlockwise). The hopping parameters
$t^{\frac{1}{2}\frac{1}{2}\frac{1}{2}}$ can have different
connections: depending on the character of the rotation of the two
octahedra the hopping parameter can increase, decrease or be
similar in amplitude to those obtained for the Sr$_2$RuO$_4$
compound. }
\begin{center}
\begin{tabular}{|l|l|l|l|l|l|l|l|l|}
\hline
\multicolumn{1}{|l|}{orbital index} & \multicolumn{8}{c|}{amplitude} \\
\hline $[l\,m\,n]$ & [000] & [100] & [010] & [110] & [200] & [020]
& [$\frac{1}{2}$\,$\frac{1}{2}$\,$\frac{1}{2}$]
 & [00p] \\
 xy-xy & -0.482 & -0.292 & -0.292 & -0.134 & -0.021 & -0.021 & 0.002/0.001 &  -0.018\\  
 yz-xy & 0 & $\pm$0.001 & $\pm$0.010 & $\pm$0.001 & $\pm$0.002 & $\pm$0.006 & 0.006/0.005/0.004 &  0 \\
 xz-xy & 0 & $\pm$0.010 & $\pm$0.001 & $\pm$0.001 & $\pm$0.006 & $\pm$0.002 & 0.006/0.005/0.004 &  0 \\
 yz-yz & -0.386 & -0.020 & -0.301 & 0.014 & 0.002 & 0.041 &  -0.023/-0.018/-0.014 &   -0.264 \\
 yz-xz & 0 & $\pm$0.061 & $\pm$0.061 & -0.013 & $\pm$0.007 & $\pm$0.007 & -0.024/-0.015/-0.006 &  0 \\
 xz-xz & -0.386 & -0.301 & -0.020 & 0.014 & 0.041 & 0.002 & -0.023/-0.018/-0.014 &  -0.264 \\
\hline
\end{tabular}
\end{center}
\label{Tab1_327}
\end{table*}
\normalsize

In a tetragonal environment the MLWFs mainly correspond with the
$t_{2g}$-atomiclike states, and this is the case for the {\sroa}.
However, the MLWFs for the {\srob} are different because the
rotation of the octahedra slightly modifies their character by
leading to a small charge redistribution following the distortion
of the orthorhombic structure. In particular, this misalignment
turns out to be more pronounced for the MLWFs which has an orbital
distribution with $d_{xy}$ symmetry. Similar conclusions have been
also reported in Ref. ~\onlinecite{Piefke11}.
The effective electronic parameters for the {\srob} based on the
t$_{2g}$-like MLWFs are reported in Table~\ref{Tab1_327}. There
are different aspects that can be noticed when comparing the
effective hopping of the {\sroa} with those of the {\srob}. The
first observation is that the nearest-neighbor hopping for the
d$_{xy}$ orbital in the distorted structural configuration is
reduced with respect to the ideal tetragonal case and matches with
that one for the d$_{xz}$ and d$_{yz}$ states. The reduction for
the d$_{xy}$ orbital can be also phrased in a Slater and Koster
scheme\cite{Slater54}. Indeed, the integral overlap
$E_{d_{xy},p_{x}}$ between the d$_{xy}$ orbital on Ru and the 2p
orbital on O reduces when there is a modification of the angle
between the Ru and O atoms. Since the effective Ru-Ru hopping
between the d$_{xy}$ orbitals $t^{100}_{xy,xy}$ depends on the
amplitude $E_{d_{xy},p_{x}}$ the
reduction expected from the Slater and Koster approach is consistent with the result obtained within the MLWFs approach.
As an indirect consequence of the change of the $xy$ nearest
neighbor hopping, the VHS moves below the Fermi level as shown in
Fig. \ref{fig:BS_327}.

Then, one can notice that the Ru-Ru hybridization via the d$_{xy}$
orbital occur both within the bilayer and also for the
second-nearest neighbor in the RuO layers. The d$_{xy}$ orbital
exhibits also non negligible overlap with the d$_{xz}$ and
d$_{yz}$ within the RuO layers. Such hybridization processes are
identically zero by symmetry in the {\sroa} system and thus
represents a relevant contribution in the determination of the
electronic structure and consequently the Fermi surface topology
for the {\srob}. Concerning the $c$-axis dispersion, the $\gamma
z$ orbitals have a larger overlap across the bilayers if compared
to the {\sroa} system.

Finally, concerning the structural analysis, looking at
Table~\ref{Tab_InterfaceVolume} we find out that the $a$ lattice
constant almost coincides with the experimental value, while the
experimental $c$ lattice constant is slightly lower than the
numerical value. Also in this case, the theoretical volume is
$\sim$0.7\% larger than the experimental one.


\small
\begin{table*}
\caption{Comparison between the experimental and theoretical
estimation of the lattice constants for the {\sroa} and the
{\srob}. The length unit is in angstrom.}
\begin{center}
\begin{tabular}{|l|l|l|l|l|l|l|}
\hline
 &\small Exp. {\sroa} ~\cite{Chmaissen98} &\small Th. {\sroa} &\small Exp. {\srob} ~\cite{Huang98} &\small Th. {\srob} &\small (Sr$_2$RuO$_4$)$_{4}$-(Sr$_3$Ru$_2$O$_7$)$_{2}$ &\small (Sr$_2$RuO$_4$)$_{3}$-(Sr$_3$Ru$_2$O$_7$)$_{3}$ \\
\hline
 $a$ & 3.862 & 3.887 & 3.873 & 3.872 & 3.881  & 3.869    \\ 
 $c$ & 12.723 & 12.650 & 20.796 & 20.968 & 46.234 & 50.554   \\ 
\hline
\end{tabular}
\end{center}
\label{Tab_InterfaceVolume}
\end{table*}
\normalsize


\subsubsection{{\sroa}-{\srob} superlattice: structural properties}

In this subsection we consider the structural properties of the
Sr$_2$RuO$_4$-Sr$_3$Ru$_2$O$_7$ superlattices. The fully relaxed
volume and the intra unit cell atomic positions have been
determined for the two superlattices, shown in Fig.
\ref{fig:strutt}(c)-(d), of the type
(Sr$_2$RuO$_4$)$_l$-(Sr$_3$Ru$_2$O$_7$)$_m$ with $(l,m)$=$(3,3)$ and
$(4,2)$, where $l$ and $m$ are the number of $n$=1 and $n$=2 unit
cells in the heterostructure. Hereafter we denote as HET42 and
HET33 the two superlattice structures with $(l,m)$=$(4,2)$ and
$(3,3)$, respectively.

As a first outcome of the structural analysis, we have compared
the lattice constants of the $n$=1 and $n$=2 bulk phases with the
ones obtained by the full relaxation of the HET42 and HET33. The
results are reported in Table~\ref{Tab_InterfaceVolume} and
include also the comparison with the experimental data for the
available compounds. When the HET42 and HET33 configurations are
considered, we find that the HET42 $a$ lattice constant has an
intermediate value between the amplitude of the {\sroa} and
{\srob} bulk phases. Otherwise, for the case of the HET33
superlattice, where the number of {\srob} cells is increased, $a$
gets further reduced if compared to the theoretical values
obtained for the pure {\sroa} and {\srob}. Such modification of
the in-plane lattice constants influences the dimension of the
unit cell along the $c$-axis by leading to an overall elongation.

\begin{figure}[!]
\centering
\includegraphics[width=0.45\textwidth, angle=0]{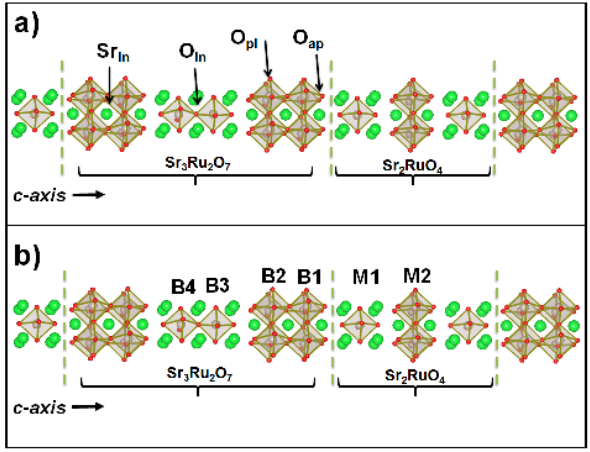}
\caption{Notation about the labelling of the atoms within the
superlattice.  O$_{pl}$ is the planar oxygen, O$_{ap}$ is the
apical oxygen, O$_{in}$ denotes the intra-bilayer oxygen, and
Sr$_{in}$ the intra-bilayer strontium. Sr$_{in}$ and O$_{in}$ are
present only in the {\srob} side of the heterostructure. The remaining strontium atoms are labelled Sr.}
\label{fig:Notation_heterostructure}
\end{figure}

The analysis of the atomic positions provides indications of the
rearrangement of the atoms within the Sr$_2$RuO$_4$ and
Sr$_3$Ru$_2$O$_7$ sides of the superlattice. From the inspection
of the Tables \ref{Tab_InterfaceDistance} and
\ref{Tab_InterfaceAngle} we are able to extract all the relevant
information concerning the distortions and rotations of the
RuO$_6$ octahedra in the superlattice. In particular, due to the
$c$-axis mismatch between the $n$=1 and $n$=2 systems the shorter
(longer) unit cell gets elongated (flattened) to optimally adapt
within the heterostructure. Consequently the Ru-O distances are
modified in such a way that the octahedra are elongated in the
Sr$_2$RuO$_4$ and flattened in the Sr$_3$Ru$_2$O$_7$ side of the
superlattice if compared to the corresponding bulk phases. Such
distortions are not uniform because the main changes occur at the
Sr$_2$RuO$_4$-Sr$_3$Ru$_2$O$_7$ interface while in the inner
layers the octahedra relax at bond lengths which match those of
the bulk systems. Assuming the notation in Fig. \ref{fig:Notation_heterostructure}a, the Ru-O$_{in}$ distance in the
Sr$_3$Ru$_2$O$_7$ does not change for the superlattice HET42 but it grows for the HET33 structure. The growth
occurs both in the inner side and at the interface of the
superlattice. However, at the interface the distance Ru-O$_{in}$ increases respect to the inner layers in both cases.
The bond lengths involving the Sr-atoms and the surrounding
oxygens are also modified in the superlattice if compared to the
bulk phases. The major changes are observed in the
Sr$_3$Ru$_2$O$_7$ side of the superlattice where the Sr-O bonds
have the tendency to increase with respect to the bulk values.

\begin{figure}[!]
\centering
\includegraphics[width=0.25\textwidth, angle=0]{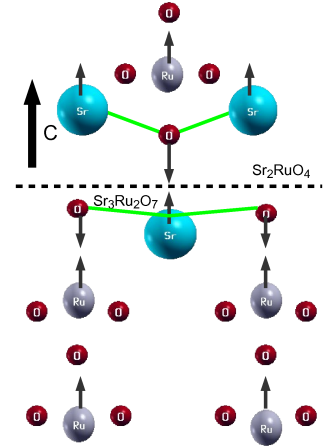}
\caption{Schematic view of the atomic rearrangement at the
interface of the Sr$_2$RuO$_4$-Sr$_3$Ru$_2$O$_7$ heterostructure.
The arrows indicate the most relevant displacements at the
interface as compared to the inner layers. The different
geometrical configuration of the Sr-O plane is shown (green line).
It is more flat for {\srob}, while there is a greater difference
between the Sr and the O atom along the $c$-axis in {\sroa}.}
\label{fig:Scheme_heterostructure}
\end{figure}

To complete the structural analysis, we have considered the
$\Delta z$ displacement along the $c$-axis of the Ru atoms with
respect to the planar oxygens, and the variation of the Ru-O-Ru
bond angles. The results are summarized in
Table~\ref{Tab_InterfaceAngle} from which we can infer two
different trends for $\Delta z$ in {\sroa} and {\srob}. Indeed, as
far as $\Delta z$ for {\sroa} is considered, we see that in the
inner unit cells of both the HET42 and HET33 structures this
quantity is zero, i. e. there is no variation of the displacement
along the $c$-axis of Ru ions compared to the pure phase.
On the other hand, at the interface a small $\Delta z$ is
produced and the Ru atom goes far from the interface as shown in Fig. \ref{fig:Scheme_heterostructure}. Concerning the {\srob}, a $\Delta z$ is already present
in the case of the bulk phase, but an enhancement of the amplitude
of $\Delta z$ is deduced at the interface.

Concerning the Ru-O-Ru bond angles there is no significant
variation in the inner layers of the {\sroa} side of the
superlattice for both the HET42 and HET33 configurations, while at
the interface the displacement along the $c$-axis of the Ru atoms
produces a reduction of the Ru-O-Ru angle. The situation is
completely different for the {\srob} side of the hybrid structure.
Indeed, a deviation from the pure phase angle is found, as well as in the
inner region, exhibits a dependence whose amplitude is related to the size
of considered superlattice.
However, the bond angle increases respect to the inner region in both cases.
The rotation angle of the octahedra, that is half of the Ru-O-Ru supplementary angle, decreases at the
interface in the {\srob} phase.

On the basis of the results above discussed, we show how the atoms
of {\sroa} and {\srob} rearrange at the interface, along the
$c$-axis. The picture that emerges is shown in Fig.
\ref{fig:Scheme_heterostructure}. The change of the atomic
positions reveals a tendency of the ions positively charged in the
ionic picture to move in the direction of the Sr$_2$RuO$_4$ side,
while the negative ones go in the opposite direction towards the
Sr$_3$Ru$_2$O$_7$ side. We notice that two bonds are strongly
modified at the interface: the distance between the ruthenium and
the apical oxygen, and the distance between the Sr ion and the
apical oxygen along the $c$-axis, the other in-plane modifications
being smaller.

\begin{table*}
\caption{Inequivalent atomic bonds length in bulk ruthenates and
for the (Sr$_2$RuO$_4$)$_4$-(Sr$_3$Ru$_2$O$_7$)$_2$ and
(Sr$_2$RuO$_4$)$_3$-(Sr$_3$Ru$_2$O$_7$)$_3$ superlattices. The
inner layers are the RuO planes which have neighbors along the
$c$-axis having the same unit cell. For instance, in Fig.
\ref{fig:Notation_heterostructure}b they are labelled as B3 and B4
for the Sr$_3$Ru$_2$O$_7$ region, and  M2 for the Sr$_2$RuO$_4$
one. The interface layers are those at the boundary between the
Sr$_2$RuO$_4$ and Sr$_3$Ru$_2$O$_7$ regions of the superlattice.
In Fig. \ref{fig:Notation_heterostructure}b they are denoted as B1
and B2 for the Sr$_3$Ru$_2$O$_7$ region, and  M1 for the
Sr$_2$RuO$_4$ one. 
There are empty spaces in the table because the corresponding values are not allowed for the given crystal structure.}
\begin{center}
\begin{tabular}{|l|l|l|l|l|l|l|}
\hline
&\tiny Exp. ~\cite{Chmaissen98,Huang98} &\tiny Th. Bulk &\tiny Inner layers HET42 &\tiny Interface layers HET42 &\tiny Inner layers HET33 &\tiny Interface layers HET33 \\
\hline
Ru-O$_{ap}$ in {\sroa}              & 2.062           &  2.059      &  2.059      &  2.069     &  2.068      &   2.076      \\
Ru-O$_{ap}$ in {\srob}              & 2.038           &  2.059      &  2.058      &  2.050     &  2.063      &   2.056      \\
\hline
Ru-O$_{pl}$ in {\sroa}              & 1.931           &  1.943      & 1.943       &  1.943     &  1.934      &   1.934      \\
Ru-O$_{pl}$ in {\srob}              & 1.956           &  1.972      & 1.977       &  1.977     &  1.973      &   1.972      \\
\hline
Ru-O$_{in}$ in {\srob}              &  2.026          &   2.045     &  2.043      &  2.045     &  2.049      &   2.052      \\
\hline
Sr-O$_{ap}$ in {\sroa} along $c$    & 2.429           &  2.433      &  2.433      &       &  2.431     &        \\
Sr-O$_{ap}$ in {\srob} along $c$  & 2.452           &  2.449      &  2.455      &       &  2.447      &        \\
\hline
Sr$^{(n=2)}$-O$_{ap}^{(n=1)}$ along $c$    &           &       &       & 2.423      &       &   2.416      \\
Sr$^{(n=1)}$-O$_{ap}^{(n=2)}$ along $c$  &            &        &        & 2.472      &        &   2.469      \\
\hline
Sr-O$_{ap}$ in {\sroa} in $ab$        &  2.738          &  2.757      &   2.757     &  2.758     &  2.745      &   2.747      \\
Sr-O$_{ap}$ in {\srob} in $ab$        &  2.744          &  2.743      &   2.755     &  2.755     &  2.739      &   2.740      \\
\hline
Sr-O$_{pl}$ in {\sroa}              & 2.688           &  2.670      &  2.671      &  2.673    &  2.676      &   2.675      \\
Sr-O$_{pl}$ in {\srob}              & 2.506/2.896     &  2.473/3.002      &  2.478/2.997      &  2.480/2.997   &  2.470/3.015      &  2.473/3.017        \\
\hline
Sr$_{in}$-O$_{in}$                   &  2.738          &  2.738      &   2.747     &  2.747     &   2.737     &   2.736     \\
\hline
Sr$_{in}$-O$_{pl}$                   & 2.607/2.986     &  2.548/3.064      &  2.556/3.062      &  2.553/3.055    &  2.545/3.074      &  2.543/3.068       \\
\hline
\end{tabular}
\end{center}
\label{Tab_InterfaceDistance}
\end{table*}
\normalsize

\small
\begin{table*}
\caption{Ru-O-Ru bond angles and displacement of Ru in the several
case studied. At the interface, the modification of the Ru-O-Ru
bond angle in {\sroa} it is due to the Ru displacement along the
$c$-axis, no rotations are found.}
\begin{center}
\begin{tabular}{|l|l|l|l|l|l|l|}
\hline
&\tiny Exp. ~\cite{Chmaissen98,Huang98} &\tiny Th. Bulk &\tiny Inner RuO layers HET42 &\tiny Interface HET42 &\tiny Inner RuO layers HET33 &\tiny Interface HET33 \\
\hline
 $\Delta z$    in {\sroa}  & 0 & 0 & 0 & 0.008 & 0 & 0.007 \\
 $\Delta z$    in {\srob}  & 0.017 & 0.033 & 0.032 & 0.039 & 0.030 & 0.037 \\
 Ru-O-Ru bond angle in {\sroa}  & 180.0$^\circ$ & 180.0$^\circ$ & 180.0$^\circ$ & 179.5$^\circ$ & 180.0$^\circ$ & 179.6$^\circ$ \\
 Ru-O-Ru bond angle in {\srob}  & 163.9$^\circ$ & 158.1$^\circ$ & 158.6$^\circ$ & 158.8$^\circ$ & 157.4$^\circ$ & 157.5$^\circ$ \\
\hline
\end{tabular}
\end{center}
\label{Tab_InterfaceAngle}
\end{table*}
\normalsize

\subsubsection{{\sroa}-{\srob} superlattice: electronic properties}

We have determined the electronic properties of the
(Sr$_2$RuO$_4$)$_4$-(Sr$_3$Ru$_2$O$_7$)$_2$ and
(Sr$_2$RuO$_4$)$_3$-(Sr$_3$Ru$_2$O$_7$)$_3$ heterostructures with
the aim to analyze the modifications of the spectra within the
superlattice and in comparison with the bulk phases as well as to
extract the interrelation between the structural changes and the
electronic dispersions. Based on the detailed analysis of the
structural properties we expect that the effective hopping and the
hybridization parameters for the energy bands at the Fermi level
are influenced both in the amplitude and in the character. Also a
rearrangement of the on-site Ru 4$d$ energies is expected to
influence the 4$d$ energy splitting due to the flattening and
elongation of the RuO$_6$ octahedra within the superlattice.

Since the unit cell is quite complex and contains many bands close
to the Fermi level we have analyzed the projected density of
states for the Ru t$_{2g}$-like orbitals as well as for the oxygen
and Sr bands at the inner and the interface layers of the
Sr$_2$RuO$_4$ and Sr$_3$Ru$_2$O$_7$ sides within the superlattice.
Furthermore, in order to extract the relevant changes induced by
the interface reconstruction and the structural distortions we
have determined the effective tight-binding Hamiltonian in the
MLWFs basis for the bands close to the Fermi level.

\begin{figure}[!]
\centering
\includegraphics[width=0.35\textwidth, angle=270]{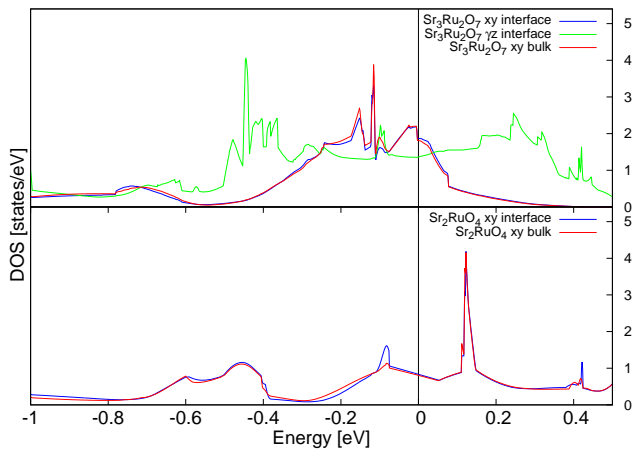}
\caption{The projected density of states for the 4d bands of the
Ru atoms placed at different layers within the HET42
heterostructure for the {\srob} (top panel) and {\sroa} (bottom
panel) sides. The blue line represents the d$_{xy}$ DOS for Ru
atoms at the interface, the red lines the DOS in the inner layers.
The Fermi energy is set to zero.}
\label{fig:DOS_Ru}
\end{figure}

Let us consider the DOS for the Ru t$_{2g}$-like bands. The
results are presented for the HET42 configuration. We have checked
that the HET33 structure does not exhibit substantial qualitative
differences. In Fig. \ref{fig:DOS_Ru} it is reported the DOS for
the projected t$_{2g}$-like bands of Ru atoms placed at the
interface and inner layers for the {\sroa} and {\srob} side of the
superlattice for an energy window close to the Fermi level where
the major changes occur. From a general view the DOS shows the
features of the bulk {\sroa} and {\srob} for instance as far as it
concerns the Van Hove like peaks below and above the Fermi level
for the d$_{xy}$ bands and the one dimensional distribution of
spectral weight for the $\gamma z$ ones. One important aspect we
notice is that, though the octahedra deformation is not uniform
within the superlattice, the DOS does not exhibit significant
shifts in energy when comparing the interface with the inner
layers. This can be addressed by considering that the change in
the crystal field and the modification of the effective bandwidth
can balance and reduce the energy shifts within the superlattice.
The analysis of the effective tight-binding parameters within the
MLWFs reveal that indeed this is the case for the {\sroa}-{\srob}
heterostructure. Slight changes are visible only for the d$_{xy}$
bands whereas in the Sr$_3$Ru$_2$O$_7$ there is a small
suppression of spectral weight below the Fermi level at the
interface compared to the inner layers, the opposite occurs above
the Fermi energy. The Ru d$_{xy}$ DOS in the {\sroa} side of the
superlattice exhibits a slight increase of the spectral weight at
energies where a redistribution and an overlap of the d$_{\gamma
z}$ bands with the d$_{xy}$ one occur in the Sr$_3$Ru$_2$O$_7$
region. No significant variations can be noticed for the $\gamma
z$ bands across the superlattice.

\begin{figure}[!]
\centering
\includegraphics[width=0.35\textwidth, angle=270]{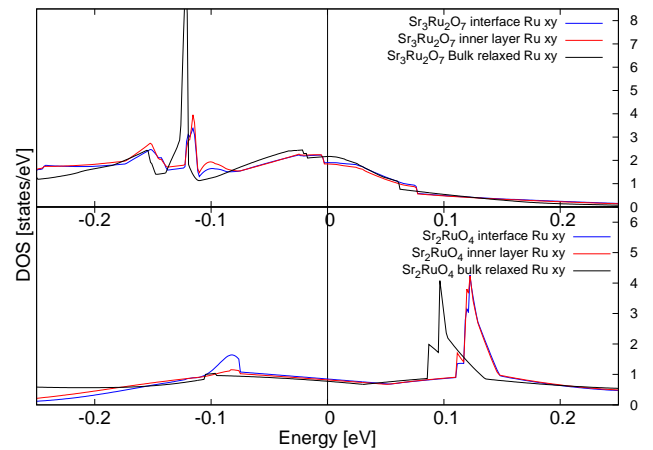}
\caption{Bulk vs heterostructure projected density of states for
the Ru $xy$ band. The Fermi energy is set to zero.
}
\label{fig:DOS_Ru_interfaccia+innerlayer+bulk_xy}
\end{figure}

More interesting is the comparison between the DOS of the
superlattice and that one of the bulk {\sroa} and {\srob} systems.
We focus on the variation of the VHS positions close to the Fermi
level for the $xy$ band. In Fig.
\ref{fig:DOS_Ru_interfaccia+innerlayer+bulk_xy} we present the
d$_{xy}$ DOS for the interface and inner layers of the HET42
superlattice for the {\sroa} and {\srob} sides in comparison with
the corresponding bulk DOS. We notice that the peak associated
with the VHS in the bulk {\srob} gets reduced and shifted towards
the Fermi level in the {\srob}. This is mainly due to the change
of the d$_{xy}$ bandwidth and of the crystal field splitting
driven by the rotation and flattening or elongation of the
octahedra. On the other hand the VHS placed above the Fermi level
in the {\sroa} is moved away from the Fermi level.

\small
\begin{table*}
\scriptsize \caption{Effective hopping parameters along the
direction $[lmn]$ and on-site energy in eV associated to the
t$_{2g}$-like Wannier functions for the ({\sroa})$_3$-({\srob})$_3$
superlattice. The connecting vector is expressed in terms of the
integer set $[l\,m\,n]$ and the lattice constants $a$ and $c$ as
$\textbf{d}=l\,a\,\textbf{x}+m\,a\,\textbf{y}+n\,c\,\textbf{z}$
for the related subsystems. The direction $00p$ connect the two
ruthenium atoms within the bilayer of the {\srob} side of the
superlattice. B1 and B2 indicate the RuO layers within the bilayer
placed at the interface with the {\sroa} and the {\srob} side of
the superlattice. B3 and B4 are the inner RuO layers in the
{\srob} part. M1 and M2 indicate the interface and the inner RuO
layers within the {\sroa} side. A schematic view of the structure
and the labels is reported in Fig.
\ref{fig:Notation_heterostructure}b.}
\begin{center}
\begin{tabular}{|c|c||c|c||c||c|c||c||c|c||c||c|c||c||c|c|}
\hline
\multicolumn{1}{|c|}{} & \multicolumn{9}{c||}{ \srob\ } & \multicolumn{1}{c||}{ Interface } & \multicolumn{5}{c|}{ \sroa\ } \\
\hline
\multicolumn{1}{|c|}{orbital} & \multicolumn{1}{c||}{ B4-B3 } & \multicolumn{2}{c||}{  B3 } & \multicolumn{1}{c||}{B3-B2} & \multicolumn{2}{c||}{ B2 } & \multicolumn{1}{c||}{ B2-B1 } & \multicolumn{2}{c||}{ B1  } & \multicolumn{1}{c||}{ B1-M1 } & \multicolumn{2}{c||}{ M1 } & \multicolumn{1}{c||}{ M1-M2 } & \multicolumn{2}{c|}{ M2 } \\
\hline $[l\,m\,n]$ & [00p] & [000] & [100] & [$\frac{1}{2}$\,$\frac{1}{2}$\,$\frac{1}{2}$] & [000] & [100] & [00p] & [000] & [100] & [$\frac{1}{2}$\,$\frac{1}{2}$\,$\frac{1}{2}$] & [000] & [100] & [$\frac{1}{2}$\,$\frac{1}{2}$\,$\frac{1}{2}$] & [000] & [100]\\
\hline
 xy-xy & -0.015 & -0.423 & -0.235 & 0.002/0.001 & -0.423 & -0.235 & -0.016 &  -0.424 & -0.237 & 0.002 & -0.490 & -0.380 & 0.002 & -0.491 & -0.381  \\
 yz-xy & 0 & 0 & $\pm$0.001 & 0.006/0.004/0.003 & 0 & $\pm$0.001 & 0 &  0 & $\pm$0.001 & 0.006/0.005 & 0 & 0 & 0.006 & 0 & 0  \\
 xz-xy & 0 & 0 & $\pm$0.016 & 0.006/0.004/0.003 & 0 & $\pm$0.016 & 0 &  0 & $\pm$0.020 & 0.006/0.005 & 0 & 0 & 0.006 & 0 & 0 \\
 yz-yz & -0.248 & -0.335 & -0.008 & -0.023/-0.015/-0.011 & -0.332 & -0.008 & -0.244 &  -0.315 & -0.008 & -0.021/-0.014 & -0.307& -0.039 & -0.020 & -0.289 & -0.039\\
 yz-xz & 0 & 0 & $\pm$0.080 & -0.025/-0.013/-0.003 & 0 & $\pm$0.079 & 0 &  0 & $\pm$0.077 & -0.020/-0.009 & 0 & 0 & -0.015 & 0 & 0\\
 xz-xz & -0.248 & -0.335 & -0.285 & -0.023/-0.015/-0.011 & -0.332 & -0.285 & -0.244 &  -0.315 & -0.277 & -0.021/-0.014 & -0.307 & -0.280 & -0.020 & -0.289 & -0.273\\
\hline
\end{tabular}
\end{center}
\label{Tab_HET33}
\end{table*}
\normalsize

To understand more deeply the differences in the electronic
properties within the superlattice and with respect to the bulk
{\sroa} and {\srob} we have determined, starting from the outcome
of the density functional analysis, the effective tight-binding
Hamiltonian in the MLWFs basis. The output of the relevant
electronic parameters connecting the Ru t$_{2g}$-like Wannier
states is given in the Table \ref{Tab_HET33} for the case of the
HET33 heterostructure. This is the more general case as it
contains inequivalent inner RuO layers for both the {\sroa} and
{\srob} subsystems. Starting from the {\srob} side of the
heterostructure we notice that the electronic parameters are quite
homogeneous within the superlattice confirming the small
variations in the DOS at the interface and in the inner RuO
layers. In particular, the small modification of the local crystal
field splitting, the bilayer splitting, and the t$_{2g}$ bandwidth
have a trend that follows the main structural changes. Indeed,
since the octahedra are flattened for the outer RuO layers at the
interface (denoted as B1 and B2 in Fig.
\ref{fig:Notation_heterostructure}b) with respect to those in the
inner layers (denoted as B3 and B4 in Fig.
\ref{fig:Notation_heterostructure}b) the energy associated to the
d$_{\gamma z}$-like Wannier states is pushed up while the d$_{xy}$
is not modified. Thus, the overall effect is to reduce the crystal
field splitting,  i.e. $\Delta_{cf}=|E_{xy}-E_{\gamma z}|$, at the
interface with respect to the {\srob} inner side of the
superlattice. On the other hand, the bilayer splitting, the
t$_{2g}$ in-plane and out-of-plane nearest-neighbour hopping are
basically uniform within the superlattice exhibiting a variation
in a energy window of 2-10 meV. In such energy range the most
significant modification is represented by the increase of the
Ru-Ru $\gamma z$ nearest-neighbour hopping when moving from the
interface to the inner layers. It is also worth noticing that the
hybridization amplitude between the $xy$ and $\gamma z$ Wannier
states is larger in the RuO layers at the interface with the
{\sroa} than in the inner ones. At this point it is also relevant
to analyze the differences of the effective tight-binding
Hamiltonian between the superlattice {\srob} side and the
corresponding bulk phase. By inspection of the Table
\ref{Tab1_327} we notice that, due to a larger tilting of the
octahedra with respect to the bulk, the in-plane $xy$ Ru-Ru
nearest-neighbor hopping is reduced in the superlattice. There
occurs also a decrease of the $\gamma z$ Ru-Ru nearest-neighbor
hopping but this is mainly driven by the flattening of the RuO$_6$
octahedra in the superlattice. The distortion of the octahedra
influences also the crystal field and the bilayer splitting. The
energy splitting $\Delta_{cf}$ is generally reduced in the
superlattice if compared to the bulk phase except for the Ru atoms
placed in the B1 outer layer of the interface bilayer (see Fig.
\ref{fig:Notation_heterostructure}b). The hybridization amplitude
between the $xy$ and $\gamma z$ Wannier states, which is a
relevant parameter in setting the differences between the {\sroa}
and {\srob} electronic structures, is doubled at the interface of
the superlattice compared to that for the {\srob} bulk phase.

Let us consider the electronic parameters for the {\sroa} bulk and
in the superlattice. Starting from the crystal field splitting we
notice that the elongation of the octahedra at the interface
pushes down the energy of the Ru $\gamma z$ states. Such change,
in turn, leads also to an increase of the Ru-Ru in-plane
nearest-neighbor hopping moving from the inner layers to the
interface ones. The remaining tight-binding parameters keep the
symmetry connections as in the {\sroa} bulk phase. Though the
presence of a structural rearrangement at the {\sroa}-{\srob}
interface leads to flattening and rotation of the octahedra there
are no extra induced hybridizations between the t$_{2g}$-like
Wannier states. The comparison of the effective tight-binding
Hamiltonian between the {\sroa} side of the superlattice and the
corresponding bulk phase shows that the in-plane $xy$ Ru-Ru
nearest-neighbor hopping is slightly reduced in the superlattice
and the same happens of the $\gamma z$ orbitals. Hence, apart from
a renormalization of the t$_{2g}$ bandwidth and a modification of
the crystal field splitting the electronic structure in the
{\sroa} keeps its qualitative features as far as it concerns, for
instance, the nesting and the presence of the VHSs.

\section{Conclusions}

In conclusion by means of first-principles density functional
theory we determined the structural and electronic properties of
different configurations of superlattices made with {\sroa} and
{\srob} unit cells. Such analysis has been then exploited to
construct the effective tight-binding Hamiltonian in the MLWFs
basis in order to compare the relevant electronic parameters that
determine the dispersions of the t$_{2g}$-like states close to the
Fermi level.

We have shown that, due to the different symmetry and size of the
$n$=1 and $n$=2 unit cells, a rearrangement of the atomic position
takes place both within the Ru-O and Sr-O layers. The RuO$_6$
octahedra at the Sr$_2$RuO$_4$-Sr$_3$Ru$_2$O$_7$ interface get
elongated in the Sr$_2$RuO$_4$ side and flattened in the
Sr$_3$Ru$_2$O$_7$ along the $c$-axis when compared to the
octahedra in the inner layers block and those in the bulk phases
as well. Another interesting feature is the observation of a
misalignment of Sr atoms with respect to the O atoms in the Sr-O
blocks at the Sr$_2$RuO$_4$-Sr$_3$Ru$_2$O$_7$ interface. The
effect is not symmetric in amplitude on the two sides of the
superlattice at the interface. Furthermore, we have demonstrated
that similar structural changes occur also in the RuO layers close
to the interface and that they influence the Ru-O-Ru angle
resulting into a less pronounced rotation of the Sr$_3$Ru$_2$O$_7$ octahedra.

Concerning the electronic structure, the analysis via the MLWFs
approach allowed to extract all the relevant features for the
hopping and on-site energies amplitude within the superlattice and
in comparison with the corresponding bulk phases. The overall view
is that the electronic parameters are quite uniform in the
superlattice exhibiting small differences between the interface
and inner Ru bands. Furthermore, the symmetry allowed hopping in
the superlattice are analogue to those in the Sr$_2$RuO$_4$ and
Sr$_3$Ru$_2$O$_7$ bulk phases. This implies that, for instance,
the nesting of the $\gamma z$ is not affected qualitatively in the
Sr$_2$RuO$_4$ side as well as the presence and the character of
the VHSs.

Let us consider the connection between the presented results and
the possible experimental consequences. The first observation is
that the resulting DOS for the Ru $xy$ band at the interface in
the Sr$_2$RuO$_4$ side turns out to be greater than that in the
bulk Sr$_2$RuO$_4$. Still, the reduction of the bandwidth of the
Ru $xy$ band in the superlattice would point towards an
enhancement of the correlations at the interface. Those aspects
both indicate a tendency to increase the superconducting critical
temperature either viewed in a Bardeen-Cooper-Schrieffer scenario
or in a correlated driven pairing. Still, the increase of the DOS
at the Fermi level and the change in the effective bandwidth can
also lead to a ferromagnetic instability within a Stoner picture
thus one cannot rule out at this level of description the
possibility that the superlattice can exhibit a ferromagnetic
transition. So far, the eutectic systems are the only available
Sr$_2$RuO$_4$-Sr$_3$Ru$_2$O$_7$ heterostructure where recent
electric transport and muons measurements in the
Sr$_2$RuO$_4$-Sr$_3$Ru$_2$O$_7$ eutectic have confirmed the
occurrence of superconductivity at a temperature that is higher
than that observed in the {\it pure} Sr$_2$RuO$_4$ with an onset
of about 2.5 K. \cite{Shiroka12} Concerning the Sr$_3$Ru$_2$O$_7$
side, we have shown that the VHS is shifted towards the Fermi
level in the superlattice if compared to the bulk phase. The
closeness of the VHS to the Fermi energy is known to lead to a
reduction of the metamagnetic critical field if analyzed in the
framework of a weakly correlated approach to the metamagnetic
instability \cite{Autieri12}. Such observation can be used to
understand the origin of the metamagnetism in the
Sr$_3$Ru$_2$O$_7$ with respect to the presence of the VHS as well
as the limits of a Stoner scenario to address the response upon an
applied magnetic field.

\begin{acknowledgments}
We acknowledge useful discussions with G. Cantele and E. Pavarini.
Carmine Autieri acknowledges financial support from "Fondazione Angelo Della Riccia".
The research leading to these results has received funding from
the EU -FP7/2007-2013 under grant agreement N. 264098 - MAMA.
\end{acknowledgments}



\end{document}